\documentclass{eccm-ecfd}

\usepackage{graphicx}
\usepackage{amsmath}
\usepackage{amsfonts}
\usepackage{amssymb}

\usepackage{wrapfig}
\usepackage{caption}
\usepackage{subcaption}
\usepackage{nicefrac}

\usepackage{lipsum}

\usepackage{capt-of}
\usepackage{booktabs}
\usepackage{varwidth}
\title{Using adjoint CFD to quantify the impact of manufacturing variations on a heavy duty turbine vane}

\author{Alexander Liefke$^{1}$, Vincent Marciniak$^{1}$ , Uwe Janoske$^{2}$, Hanno Gottschalk$^{2}$}

\heading{A. Liefke, V. Marciniak, U. Janoske and H. Gottschalk}

\address{$^{1}$ Siemens AG Power and Gas, M{\"u}ülheim an der Ruhr , Germany,\vspace{-10pt}
	\and
	\{alexander.liefke, vincent.marciniak\}@siemens.com\vspace{-10pt}
	\and
	$^{2}$ Bergische Universit{\"a}t Wuppertal, Wuppertal, Germany\vspace{-10pt}
	\and
	\{uwe.janoske, hanno.gottschalk\}@uni-wuppertal.de}\vspace{-10pt}


\begin{document}

\thispagestyle{empty}
\section{Introduction}
Heavy duty gas turbines often exhibit a scatter in aerodynamic performance. This scatter is caused by, among other things, geometry variations induced by the manufacturing process. To take into account the scatter in performance, margin adaptation factors are applied after the design. A more accurate modeling of the impact of manufacturing variations (MVs) would therefore lead to a reduction in performance variation and thus higher overall efficiency, if included in the blade design process.

The analysis of MVs consists of two parts: the modeling of MVs and their impact evaluation. A first approach to investigate the impact of MVs is presented by Garzon and Darmofal \cite{GD}. The authors use low-fidelity coordinate measuring machine (CMM) scans of a compressor blade to model the MVs and evaluate their impact with a 2D boundary layer flow solver coupled with Monte Carlo simulations. Duffner \cite{Duf} also adopts this method for turbine vanes.
A high-fidelity approach is presented by Lange et al. \cite{LVVSJG} who use optical white-light scans of compressor blades, while using RANS in conjunction with a Monte Carlo simulation to quantify the impact.

In order to circumvent the computational cost of RANS evaluations, Giebmanns et al. \cite{GBFS}  employ an adjoint solver to investigate the impact of leading edge geometry variations on Quasi-3D compressor blade sections.
A similar adjoint approach, for the impact analysis of MVs on a gas turbine blade is used by Zamboni et al. \cite{ZBB}, who use one CMM measurement to assess the impact on aerodynamic performance.
Yang et al. \cite{YXMHLL} also employ an adjoint solver to study the impact of MVs on a multistage steam turbine assuming Gaussian distributed blade thickness.

The main limitation of the adjoint method is the linearization of the governing Navier-Stokes equations, while assuming that the impact of MVs is small enough to be regarded as linear.
These studies, however, focus solely on the application of adjoint methodology for impact assessment without a thorough investigation with real MVs. 
The main aim of this paper is therefore to demonstrate that the impact of MVs on turbine blades can be considered linear and can be evaluated by the adjoint method.

The first part of this work details how the adjoint method is used to evaluate the impact of MVs, introducing an adjoint-based process toolchain.
For the second part, the limitation and capabilities of two adjoint solvers are investigated, using a public subsonic 1.5-stage axial turbine test case \cite{Gal}.
In the third part, the adjoint toolchain is applied to an industrial turbine vane for which 102 white light scans are analyzed. 
The MVs are directly modeled by means of mesh morphing and are used to validate the adjoint evaluation approach with finite differences.
\section{Impact Evaluation with Adjoint}
\begin{figure}[tbp]
	\centering
	\includegraphics[width=0.95\linewidth]{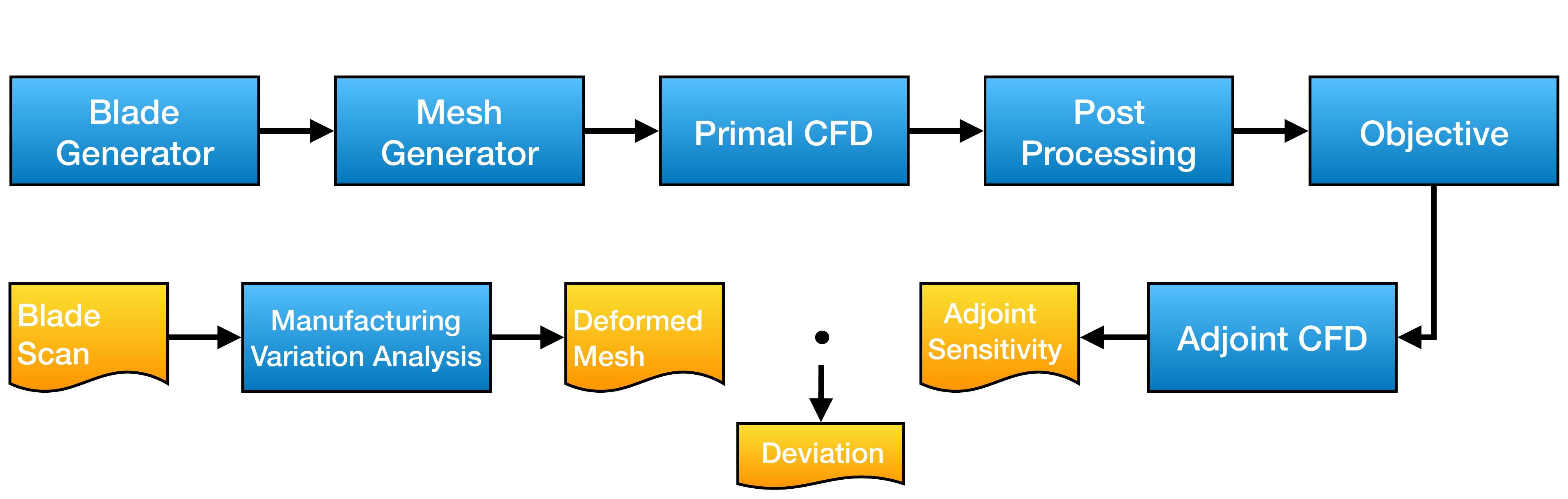}
	\caption[]{Adjoint-based Toolchain for Impact Evaluation}
	\label{fig:adjointToolChain}
\end{figure}
The impact evaluation of MVs with an adjoint CFD solver requires three steps:
\begin{itemize}
	\vspace{-5pt}
	\item perform a primal CFD evaluation of the baseline geometry\vspace{-8pt}
	\item compute one adjoint solution for each objective function \vspace{-8pt}
	\item provide deformed mesh files representing the manufacturing variations
\end{itemize}
The key process steps for the impact analysis are outlined in Figure \ref{fig:adjointToolChain}.
At first the baseline geometry is evaluated with a steady RANS computation. Then the adjoint computation with regard to a chosen objective is carried out. To evaluate the impact of MVs, the adjoint sensitivities are multiplied with a surface deformation vector field, which is based on the baseline and the deformed mesh surface coordinates. The impact of MVs is thus expressed in form of an objective gradient.

\subsection{Computational Methods}
In this work, the turbomachinery CFD suite Trace 9.1 developed at the German Aerospace Center (DLR) is used. The baseline geometry is generated with an in-house blade generator, while the mesh generation is performed using Autogrid 5. 
The turbine vane blade scans are analyzed as detailed in section  \ref{sec:manufacturingVarialionAnalysis} and use Trace Prep \cite{VFK}, the preprocessor of Trace, to perform a mesh morphing of the blade surface to create the deformed meshes.
The adjoint sensitivity is then calculated with AdjointTrace. Currently two different versions of AdjointTrace are available: a hand-derived (HD) and an algorithmic-differentiated (AD) version.

The HD version is a discrete adjoint solver, which applies manual differentiations and finite differences to compute the finite volume cell flux across each cell \cite{FKN}. For the computation of the adjoint HD solution the constant eddy viscosity (CEV) assumption is applied, which means that the turbulent quantities of the primal computation are frozen \cite{FABKW}.

The AD version is also a discrete adjoint flow solver, based on reverse mode differentiation of the primal solver \cite{BSFMNSG}. The advantage of AD is that the complete CFD code is differentiated including turbulence models. Furthermore, the adjoint solver inherits the same convergence behavior as the primal computation \cite{SOGBFMN}. The main disadvantage, however, is the high demand in memory, which compared to the primal computation has a factor of at least eight \cite{SOGBFMN}.

\section{Evaluation of Adjoint CFD Solvers}
In order to compare the two adjoint solvers, a NACA-like parametrization is applied to the rotor of an axial 1.5-stage turbine to validate the adjoint-generated gradients with finite differences. 
\subsection{Testcase Description}
\begin{figure}[b]
	\begin{minipage}{0.39\textwidth}
		\centering
		\resizebox{\linewidth}{!}{%
		\begin{tabular}{|l|c|c|c|}
			\hline 
			& Stator 1 & Rotor 1 & Stator 2 \\ 
			\hline \hline
			Tip Diameter [mm] & 600 & 600 & 600 \\ 
			\hline 
			Hub Diameter [mm] & 490 & 490 & 490 \\ 
			\hline 
			Rotational Speed [rpm] & - & 3500 & - \\ 
			\hline 
			Number of Blades [-] & 36 & 41 & 36 \\ 
			\hline 
			Radial Gap [mm] & - & 0.7 & - \\ 
			\hline \hline
			Mesh Nodes & 60,843 & 86,193 & 63,495 \\ 
			\hline 
			Radial Nodes & 17 & 25 & 17 \\ 
			\hline 
		\end{tabular}}
		\subcaption{Turbine and mesh data}
	\end{minipage}
	\hfill
		\begin{minipage}{0.59\textwidth}
		\centering
		\includegraphics[width=\linewidth]{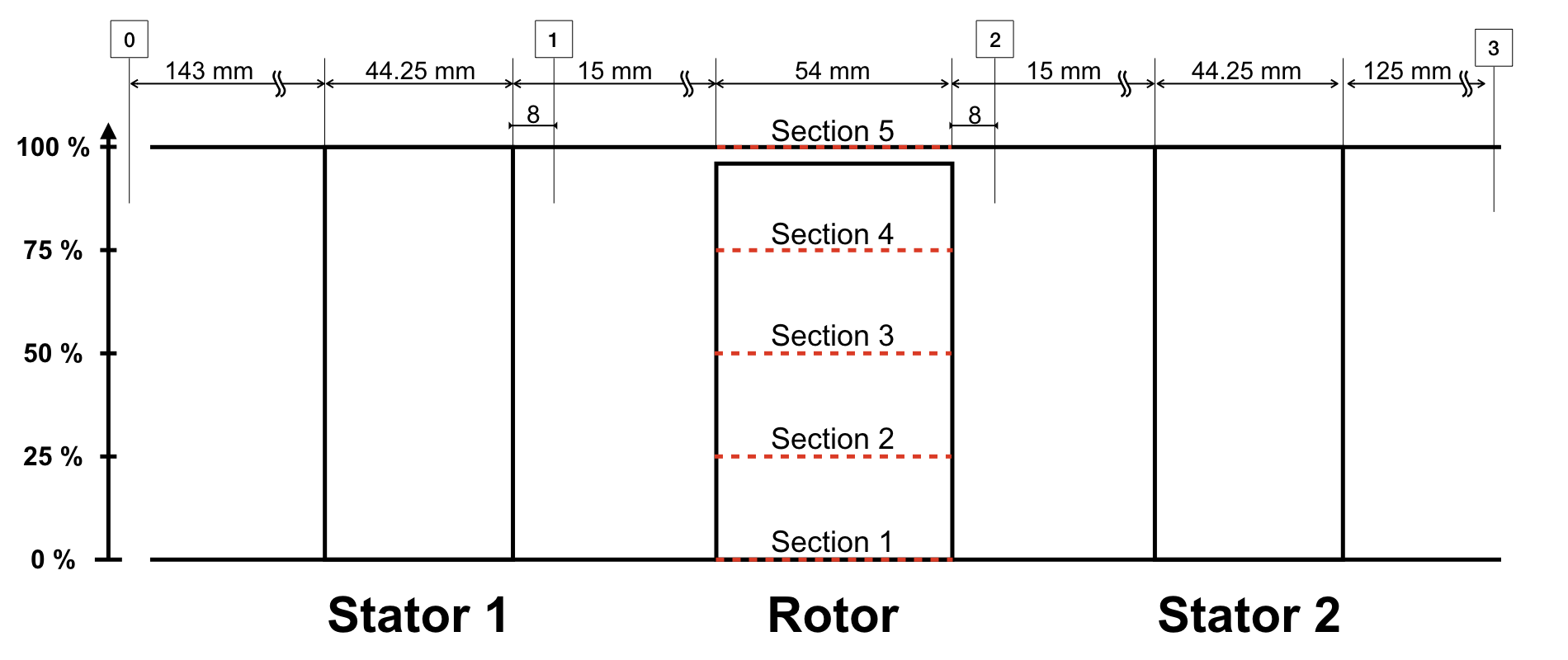}
		\subcaption{2D schematic not in scale}
	\end{minipage}
	\caption{Turbine schematic and design data}
	\label{fig:aachenGeometry}
\end{figure}
The subsonic 1.5-stage axial flow turbine consists of a stator-rotor-stator configuration and the measurements on the test case "Aachen Turbine" were carried out at the Institute of Jet Propulsion and Turbomachinery at RWTH Aachen, Germany. The turbine is used as general CFD validation test case and has been extensively studied numerically and experimentally. Figure \ref{fig:aachenGeometry} lists the basic turbine parameters and shows a two-dimensional blade schematic. Both stators and the rotor are prismatic. The computational domain is setup using the steady measurement configuration data with an average mass flow around 8 kg/s as described in the ERCOFTAC-Testcase 6 \cite{Gal}.

\subsection{Parameterization}
Only the blade geometry of the rotor blade is parameterized, using the center of gravity as stacking axis. 
The turbine blade parameters with the highest sensitivity in regard to MVs, as shown by Scharfenstein et al. \cite{SHVVM}, are: stagger angle, maximum blade thickness and maximum blade thickness curvature.

\begin{figure}[b!]
	\centering
	\begin{minipage}{0.48\textwidth}
		\centering
		\includegraphics[width=\linewidth]{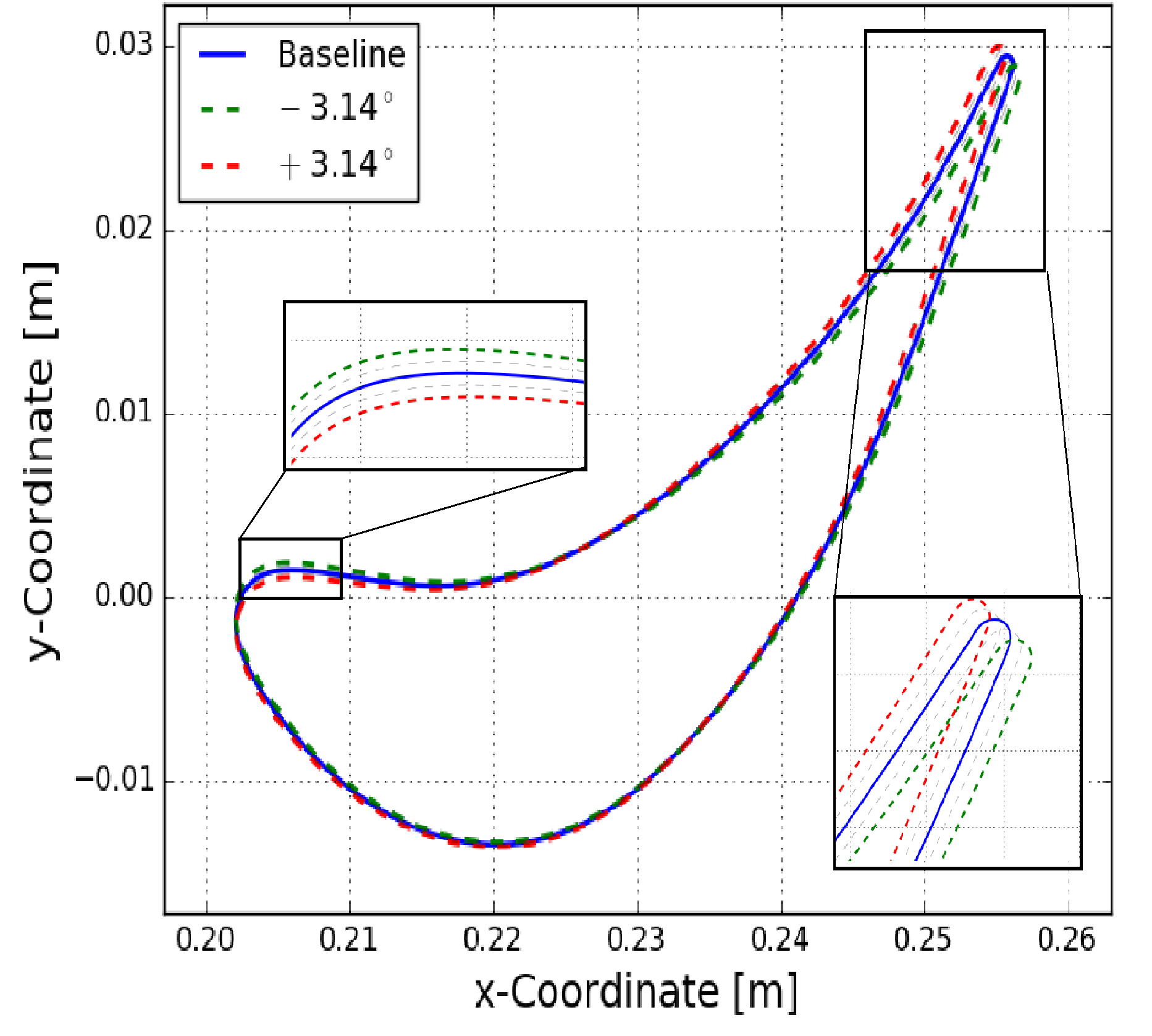}
		\subcaption{Stagger Angle}
		\label{fig:parameterDeformationStagger}
	\end{minipage}
	\hfill
	\begin{minipage}{0.48\textwidth}
		\centering
		\includegraphics[width=\linewidth]{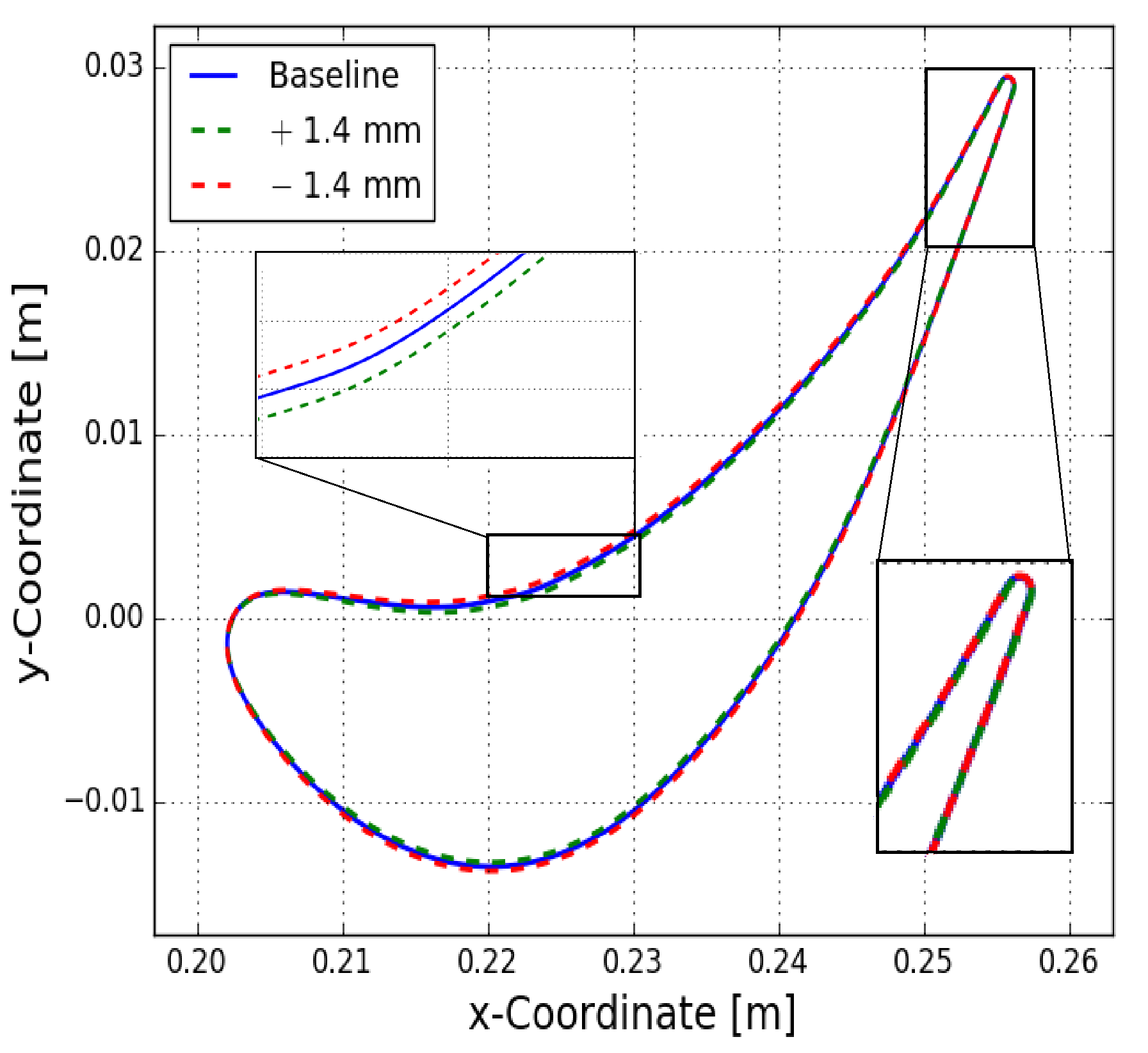}
		\subcaption{Maximum Blade Thickness}
		\label{fig:parameterDeformationThickness}
	\end{minipage}
	\caption{Effect of parameter deformation on rotor blade at 50\% span}
\end{figure}

Therefore the parameters stagger angle $\alpha$ and maximum blade thickness $D$ are chosen for a parameter study. The parameters are varied each individually on five different rotor blade sections at 0, 25, 50, 75 and 100\% span, which are highlighted in Figure \ref{fig:aachenGeometry}. For each section the parameters are changed as defined in Equation \ref{eq:parameterSteps}. This results in six  variations per parameter, twelve per section and 60 deformations in total. 

\begin{align}
\centering
	\begin{array}{c}
		\alpha = \alpha_{\text{Baseline}} \cdot \xi \\
		D = D_{\text{Baseline}} \cdot \xi\\
	\end{array}
	\hspace{10pt}with \hspace{5pt} \xi & \in \{0.90, 0.95, 0.98, 1.02, 1.05, 1.10\} 
	\label{eq:parameterSteps}
\end{align}
An example of the parameter variation magnitude for $\pm$10\% is shown in Figure \ref{fig:parameterDeformationStagger} and \ref{fig:parameterDeformationThickness}. The stagger angle is varied up to 3.14$^\circ$, while the maximum blade thickness is changed up to 1.4 mm. 
The surface between each section is interpolated with a quadratic polynomial based on the surface parameters.
\subsection{CFD Setup}
For the inlet condition at the axial plane 0, as shown in Figure \ref{fig:aachenGeometry}, a radial total pressure and temperature distribution as well as a measured inlet flow angle is applied. For the outlet condition a radial pressure distribution is set at axial plane 3. In order to couple stationary and rotating blade rows a mixing plane approach is used. The fluid is assumed to behave as an ideal gas with adiabatic walls. For the flow field of the "Aachen Turbine" transitions occurs on the first stator suction side and on the rotor suction side as reported by Restemeier et al. \cite{RJGG}. 

As noticed by the authors \cite{RJGG}, the Reynolds number is low enough so that transition is expected. This has been confirmed by a computation using transition modeling. However, for stationary gas turbines the Reynolds number is so high that transition can be neglected. Therefore in the present work only fully turbulent computations are carried out with the $k$-$\omega$ turbulence model \cite{Wil} in conjunction with Kato-Launder's \cite{Kat} stagnation point anomaly fix.

The computational grid is very coarse with 210,531 nodes using wall-functions with an average $y^+ \approx 30$. 
A mesh study is performed with an additional medium and fine mesh with 810,835 and 1,287,975 nodes; confirming a difference in mass flow of below 0.01\% between the coarse and fine mesh as well as a difference in isentropic efficiency of 0.006 points. The coarse grid is therefore deemed acceptable.
\subsection{Validation of Adjoint Trace}
The AdjointTrace gradients are validated using finite differences, comparing the objectives mass flow $\stackrel{.}{m}$ and isentropic (is) efficiency $\eta$, which is defined in Equation \ref{eq:efficiency}, using the total enthalpy $\text{H}$ at the inlet and outlet. The finite differences $\Delta F_{\text{nonlinear}}$ are calculated from the baseline primal RANS calculation and a RANS computation of the deformed geometry, as specified in Equation \ref{eq:deltaFnonlinear}.
\begin{equation}
\eta =\dfrac{\text{H}_{02} - \text{H}_{01}}{\text{H}_{02,is} - \text{H}_{01}}
\label{eq:efficiency} 
\end{equation}
\begin{equation}
	\Delta F_{\text{nonlinear}} = F(\alpha, D) - F_{\text{nonlinear}} \hspace{15pt} \text{with} \hspace{10pt} F \in \{\stackrel{.}{m}, \eta \}
	\label{eq:deltaFnonlinear} 
\end{equation}
\subsubsection{Primal and Adjoint Convergence}
The L1-Residual of the primal RANS computation drops in 2,000 time steps by six orders of magnitude in double precision and has a maximum residual of around $10^{-4}$.
For the AD AdjointTrace version the residual drops by five magnitudes for both objectives in 2,500 time steps. The convergence of the HD AdjointTrace version stagnates after 300 time steps and has a magnitude reduction of around 2.5. This, however, is insufficient to be considered converged. This is acknowledged and the comparison is continued to compare the suitability of both versions for an industrial application. 
\subsubsection{Comparison of Adjoint Results}
To validate the accuracy of the adjoint gradients, the relative gradient deviation is calculated as shown in Equation \ref{eq:deltaError}.

\begin{equation}
\centering
\frac{{\Delta F}_{\text{Nonlinear}}-{\Delta F}_{\text{Adjoint}}}{\Delta {F}_{\text{Nonlinear}}}\cdot 100 \hspace{15pt}\text{with} \hspace{15pt}F \epsilon \{\stackrel{.}{m}, \eta \}
\label{eq:deltaError}
\end{equation}
\subsubsection*{Effect on Mass Flow}
\begin{figure}[th]
	\vspace{-10pt}
	\centering
	\begin{minipage}{0.49\textwidth}
		\centering
		\includegraphics[width=1.\linewidth]{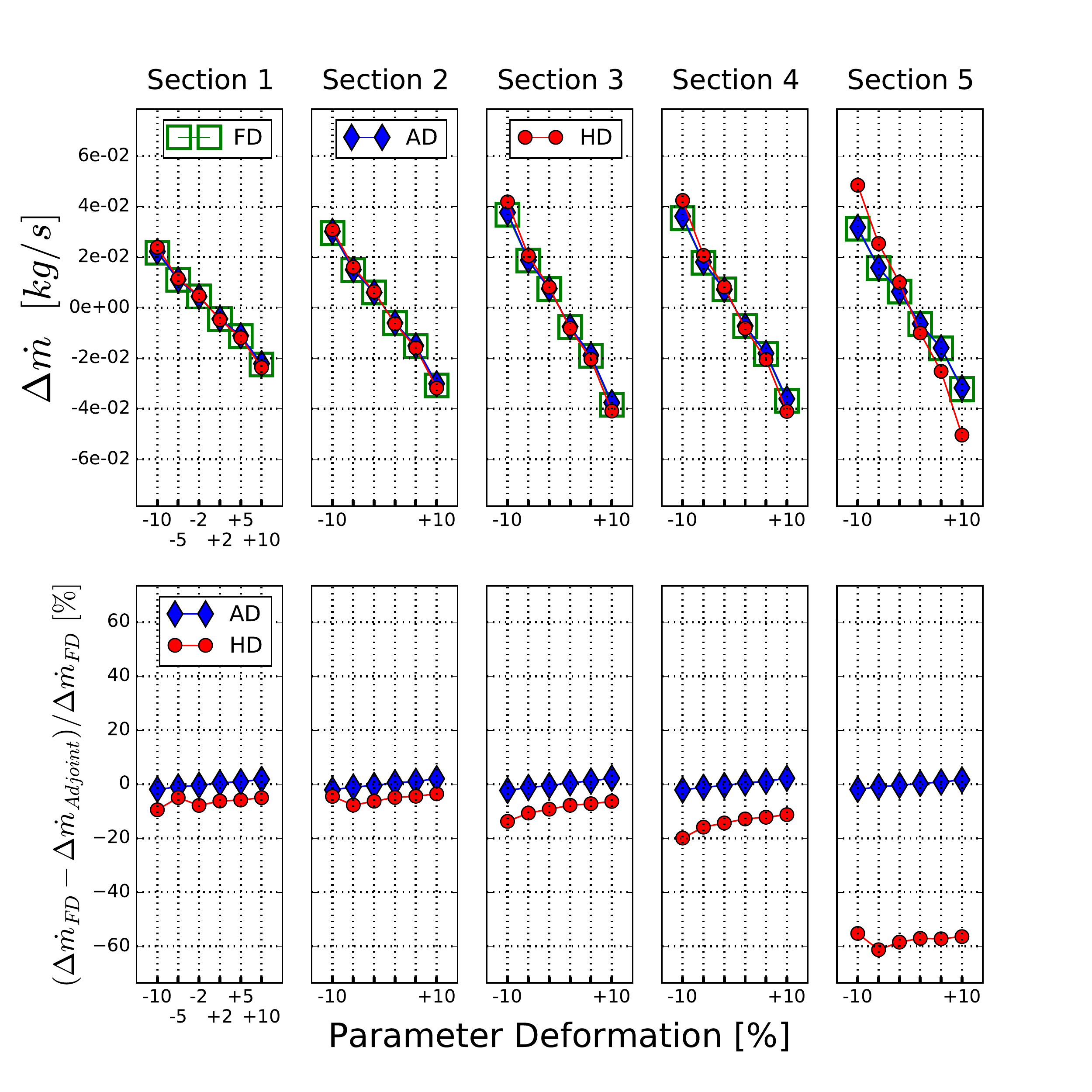}
		\subcaption{Stagger Variation}
		\label{fig:aachenMassflowStagger}
	\end{minipage}
	\hfill
	\begin{minipage}{0.49\textwidth}
		\centering
		\includegraphics[width=1.0\linewidth]{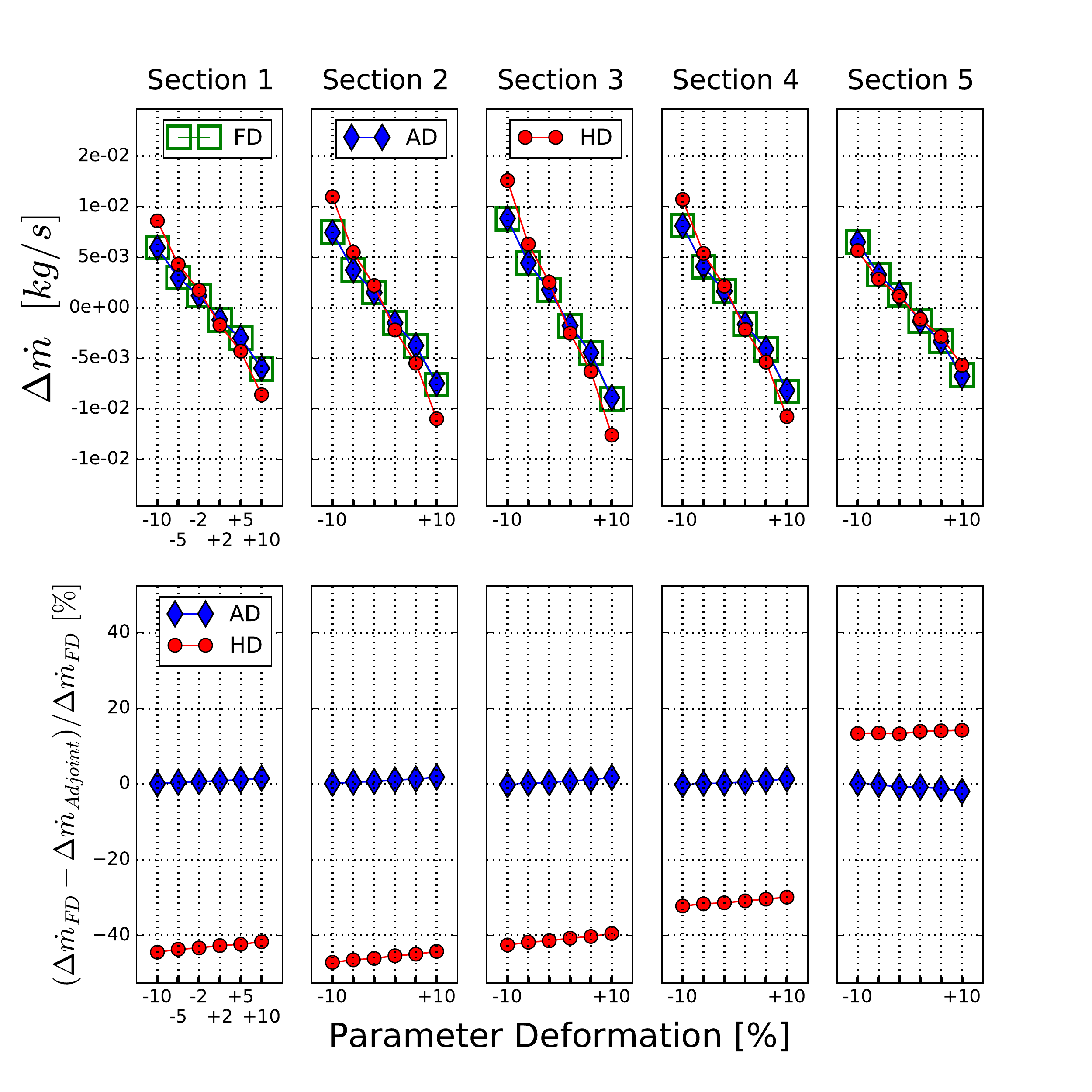}
		\subcaption{Blade Thickness Variation}
		\label{fig:aachenMassFLowThickness}
	\end{minipage}
	\caption{Impact on Mass Flow}
	\vspace{-15pt}
\end{figure}
Figure \ref{fig:aachenMassflowStagger} and \ref{fig:aachenMassFLowThickness} show the mass flow gradients for a stagger and blade thickness variation for all five sections of the AachenTurbine rotor blade. The upper Figure quantifies the absolute difference in mass flow with regard to the baseline, showing the finite difference (FD) gradients as well as the AD and HD adjoint gradients. In the lower part of the Figure, the relative deviation of the adjoint gradients in regard to finite differences is displayed in percent. 

The mass flow impact for a change in stagger angle for all five sections shows a very good agreement for the AD gradients, with a maximum deviation of $\pm$ 2 \%. The HD gradients, on the contrary, show a higher deviations of around -10\% for section one to four and around -60\% deviation for section five.

For the blade thickness variation, the adjoint AD gradients also agree very well with the finite differences, showing a deviation of $\pm$ 1\% for section one to four, while section five varies from +0.3\% to -1.4\%. The HD gradients have for the blade thickness variation a higher deviation, compared to the AD gradients, leading to deviations for section one to four of around -40\%. In contrast, the deviation for the variation at section 5 is only around +15\%.
\subsubsection*{Effect on Isentropic Efficiency}
\begin{figure}[t!]
		\vspace{-5pt}
	\centering
	\begin{minipage}{0.49\textwidth}
		\centering
		\includegraphics[width=1.\linewidth]{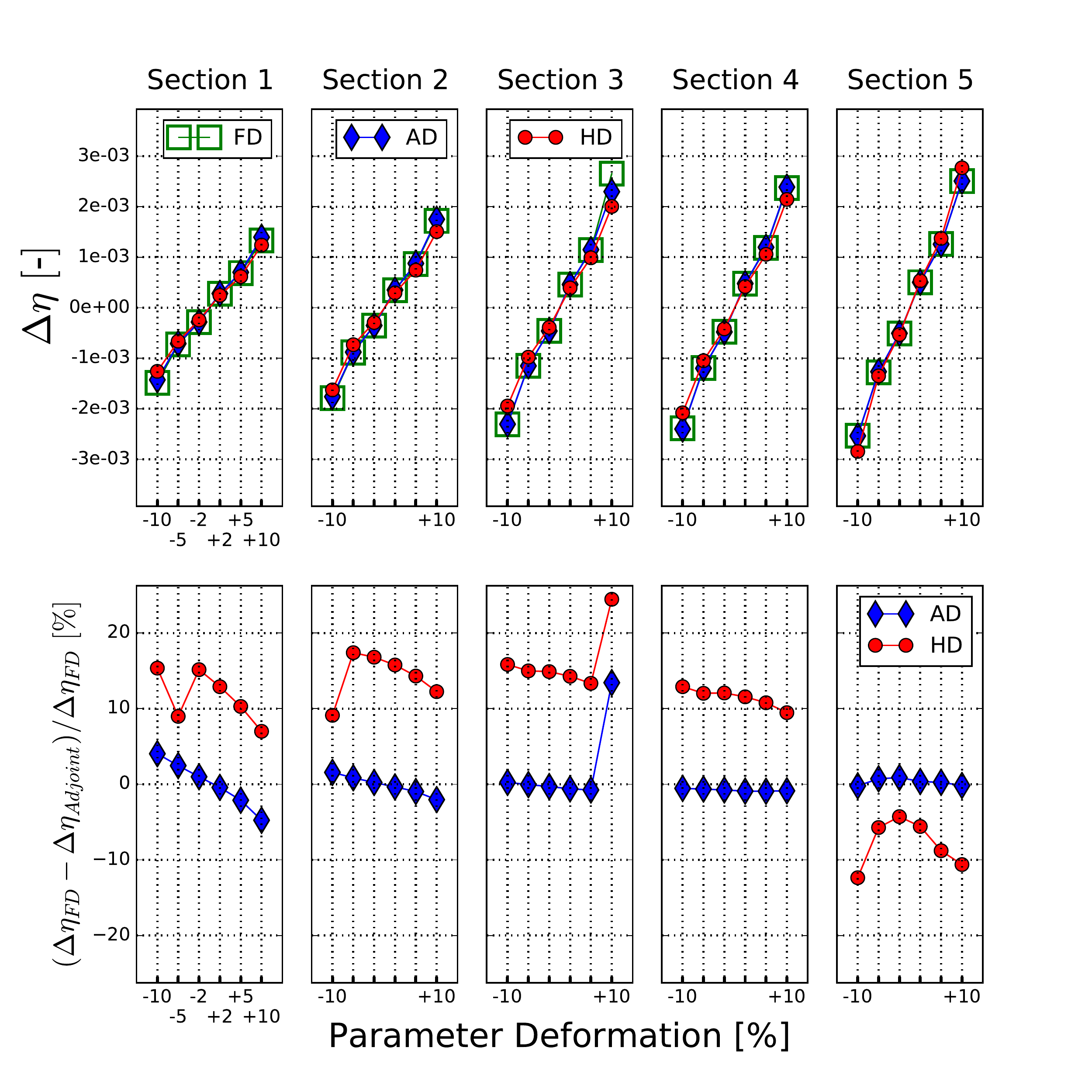}
		\subcaption{Stagger Deformation}
		\label{fig:aachenEfficiencyStagger}
	\end{minipage}
	\hfill
	\begin{minipage}{0.49\textwidth}
		\centering
		\includegraphics[width=1.00\linewidth]{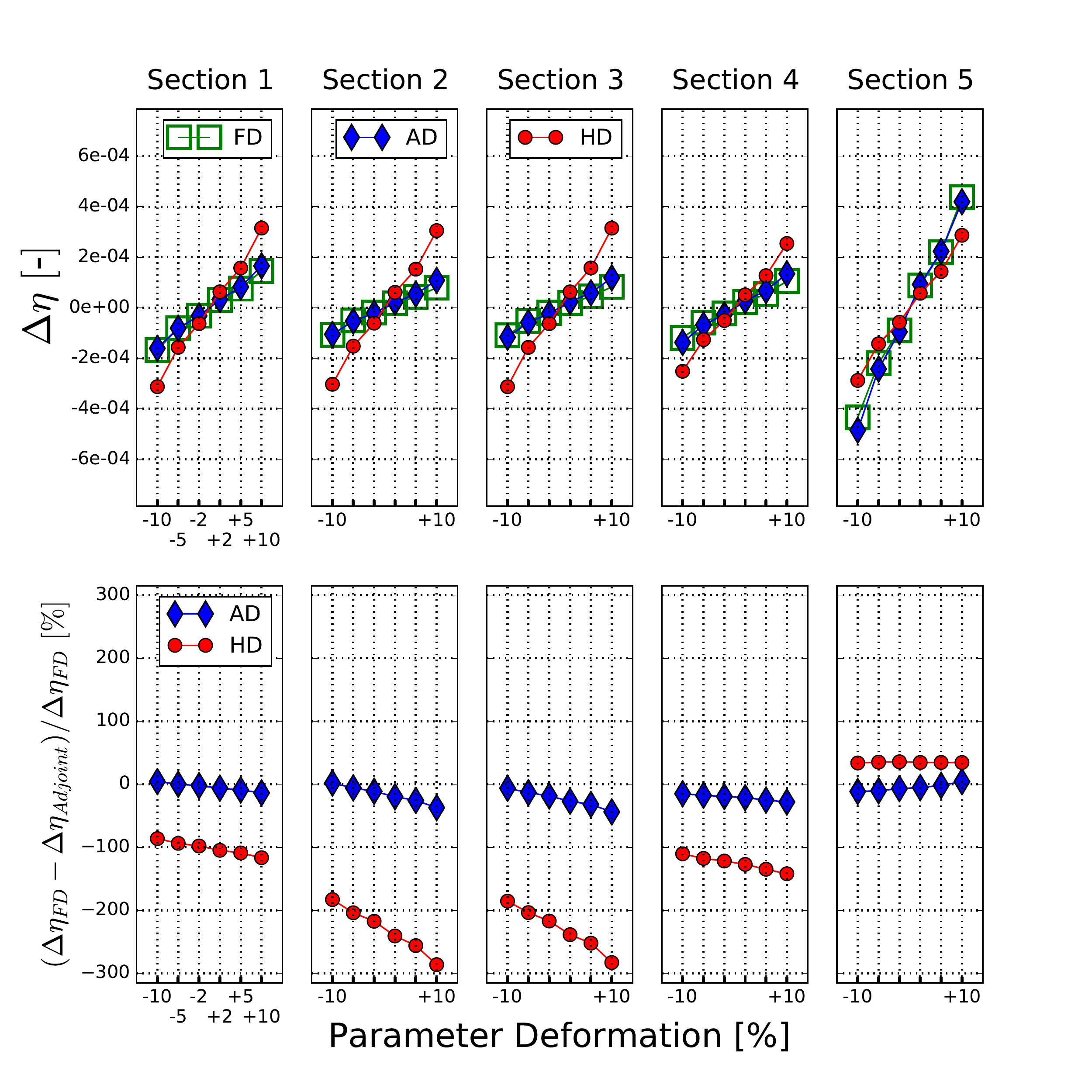}
		\subcaption{Blade Thickness Deformation}
		\label{fig:aachenEfficiencyThickness}
	\end{minipage}
	\caption{Impact on Isentropic Efficiency}
		\vspace{-15pt}
\end{figure}

In Figure \ref{fig:aachenEfficiencyStagger} the impact on the isentropic efficiency for a variation in stagger is displayed. Compared to the mass flow objective, the deviations of the AD gradients have a higher deviation of around $\pm$ 5\% for all five sections, except for a 10\% stagger variation at 50\% span. The HD sensitivities show a deviation of around $+$12\% for the first four sections and around $-$8\% for section five.

In comparison to the stagger variation, the relative blade thickness gradient deviation is one order of magnitude higher as highlighted by Figure \ref{fig:aachenEfficiencyThickness}.
At section one, the variation ranges from +4\% to -13\% for the AD gradients. For the sections two, three and four the deviations vary between +2\% and -43\%, while for section five the relative deviations are between -11\% and +4\%. On the other side, the HD gradients depict an even greater deviation, ranging from $+$35\% at section five to $-$300\% for section two and three.
\subsubsection*{Discussions}
One of the reasons for the high deviation of the HD gradients is the insufficient order in magnitude reduction of 2.5. However, the application of the AdjointTrace HD solver by Giebmanns et al. \cite{GBFS} also showed a maximum relative gradient deviation of around 110\% for a case with a residual magnitude reduction of four. Considering that a further reduction in residual could not be achieved, this also indicates a limitation of the HD AdjointTrace version.

From the results it can further be concluded that the adjoint gradients are more sensitive to a variation in blade thickness, than stagger angle.
This might be explained by the variation in boundary layer loss, caused by the change in blade surface curvature and flow acceleration on the rotor blade. Furthermore the isentropic efficiency objective has a higher relative gradient deviation compared to the mass flow gradients. The results additionally show that the isentropic efficiency sensitivity, for a thickness variation near midspan, is slightly overestimated. This might be caused by the relatively coarse mesh in radial direction, cf Figure \ref{fig:aachenGeometry}a.

From the parameter study, it can be concluded that the AD AdjointTrace version is better suited for an industrial application, introducing smaller gradient deviations and thus providing more accuracy for an impact analysis with adjoint. Therefore only the AD AdjointTrace version is used in the remaining part of this work. 

\section{Application to Heavy Duty Turbine Vane}
The test case is a first row turbine vane close to a maximum power operating point. 
In order to preserve the proprietary informations of the blade geometry, only scaled and normalized values are shown.
In total 102 turbine vane blades are analyzed using 3D optical white light scans, which have a resolution accuracy of between 10-35 $\mu m$. This is at least one order of magnitude below the turbine blade surface tolerances and therefore sufficiently accurate to model the MVs. 

Turbine blade casting involves manual production steps; after casting, the blade surface is polished and grounded with a machine by hand. 

Previous studies have used CMM or white light scans to collect MVs data, sampling the data by principle component analysis \cite{Gal} or through a parametrization approach \cite{LVVSJG}. The main limitation of both approaches is the reduction to a specific number of modes or sections to model the MVs. This paper therefore uses a direct mesh morphing approach to reduce the MVs modeling error.

The first part of this section describes the applied method to quantify the MVs. In the second part, the adjoint mesh sensitivities are analyzed. The third part quantifies the difference between the adjoint and finite difference gradients.

\subsection{Manufacturing Variation Analysis}
\label{sec:manufacturingVarialionAnalysis}

\begin{figure}[tb]
	\vspace{-10pt}
	\centering
	\includegraphics[width=1\linewidth]{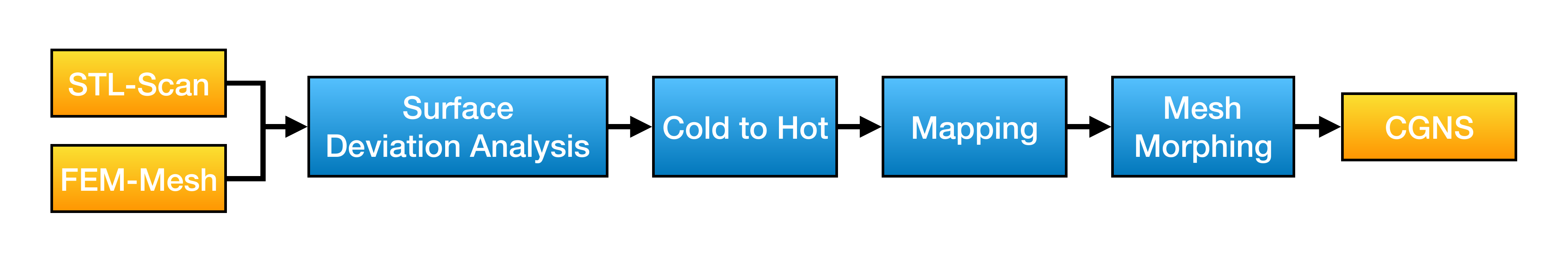}
	\caption[]{Manufacturing Variation Analysis Steps}
	\label{fig:mvaSteps}
	\vspace{-10pt}
\end{figure}

The manufacturing variation analysis steps are shown in Figure \ref{fig:mvaSteps}.
The analysis starts with two inputs: an optical scan of the manufactured blade, which stores the blade surface geometry in an STL file format, and a two-dimensional Finite Element Model (FEM) surface mesh file, from the baseline geometry. These files are used for the surface deviation analysis, quantifying the manufacturing variation by using the distance in normal direction between FEM node and STL surface. 

For the comparison of the "hot" CFD baseline geometry with the "cold" manufactured blade scans, the "cold" FEM model needs to be transformed into a "hot" state, representing the geometry deformation under hot working conditions. A cold to hot (C2H) deformation vector field is therefore applied to the FEM nodes to transform the geometry into a hot state. This assumes that the deformed surface nodes have a minor impact on the C2H deformation vector field. To prove this assumption the structural and thermal analysis for the individual morphed FEM meshes were computed. Thereby, showing that the introduced error is below 1\% for the individual node C2H vector magnitude, for 98\% of all nodes.

Next a bilinear interpolation method is used to map the deviation values from each node of the unstructured FEM model, onto the nodes of the structured CGNS. The deviation vector for each CGNS node is then used as input for the mesh morphing tool Trace Prep. Prep uses an elliptic mesh deformation algorithm as detailed in Voigt et al. \cite{VFK} and is applied to produce 102 deformed mesh geometries.

The applied mesh morphing, however, has lead to negative cell volumes near hub and shroud, which are caused by the turbine vane fillets mesh cells near the end-wall. Therefore the mesh morphing is only applied between 1-99\% span. The deformations near the hub and shroud end-wall are thus zero.
\subsection{Adjoint Results}
This part analyses the adjoint results qualitatively, before quantifying the impact of the MVs and concludes with the computational cost of adjoint.

\subsubsection{Computational Setup}
The computational grid consists of around 480,000 nodes with the dimensionless wall distance of $y^{+}\approx30$. On all surfaces wall functions are applied with adiabatic walls as well as fillets near the hub and shroud.

For the turbine vane impact assessment two objectives are evaluated: mass flow and pressure loss coefficient. The pressure loss coefficient is defined in Equation \ref{eq:pressureLoss} and is used to quantify the losses of the vane, using the total pressure at the vane inlet $\text{P}_{01}$ and outlet $\text{P}_{02}$ in relation to the static pressure $\text{p}_{02}$.
\begin{align}
\centering
\text{Y} = &\frac{\text{P}_{01} - \text{P}_{02}}{\text{P}_{02} - \text{p}_{02}}
\label{eq:pressureLoss}
\end{align}

\begin{figure}[b!]
	\vspace{-10pt}
	\centering
	\includegraphics[width=1.0\linewidth]{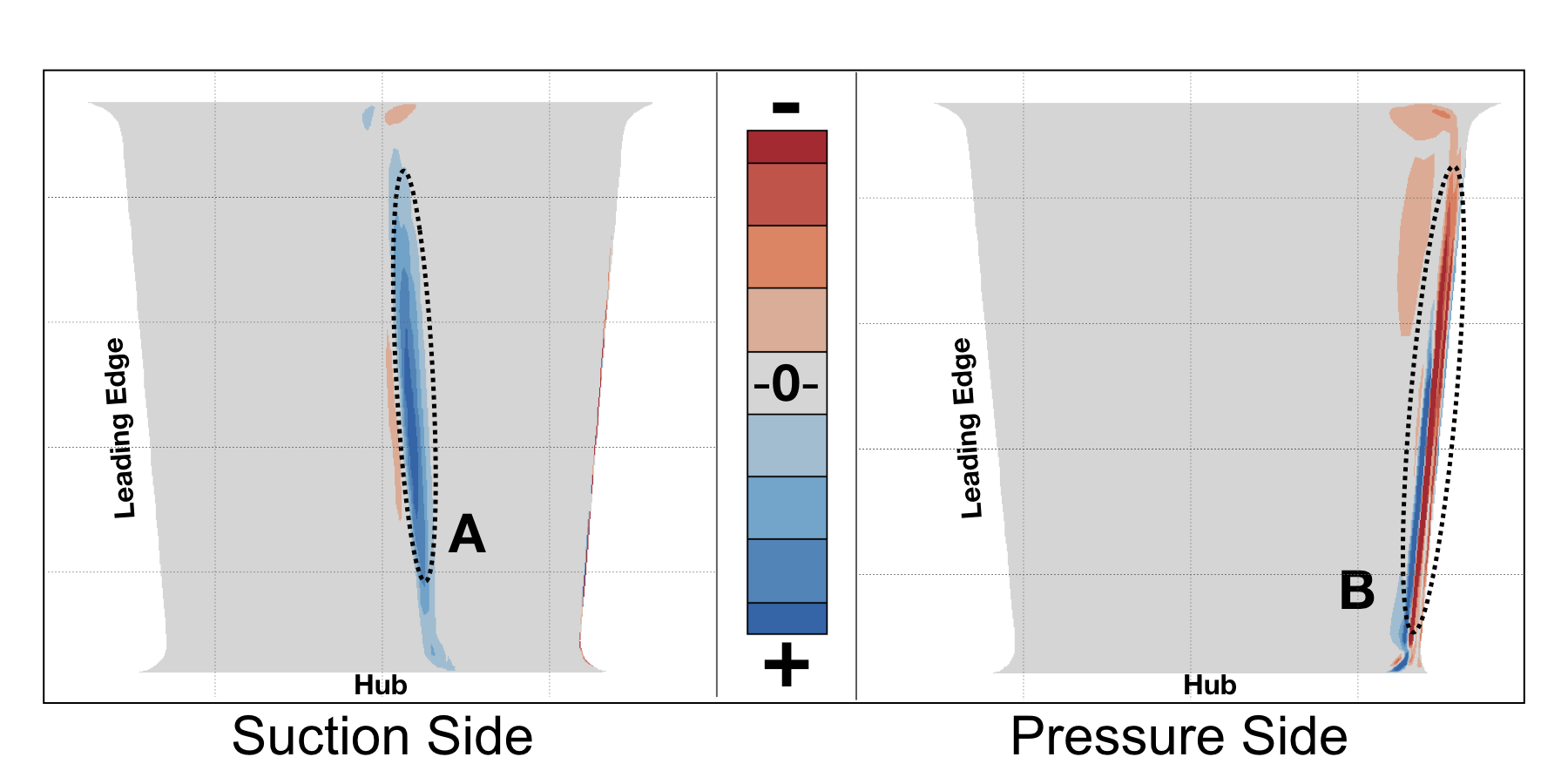}
	\vspace{-20pt}
	\caption{Mass flow surface sensitivities. Scaled suction and pressure side. Red/blue color highlighting surface outward/inward movement in normal direction. Plus/minus sign indicating increase/decrease in mass flow}
	\label{fig:surfaceSenseMassFlow}
	\vspace{-10pt}
\end{figure}
\subsubsection{Adjoint Surface Sensitivity}
The outcome of an AdjointTrace computation are sensitivities with regard to each mesh node of the complete domain. To obtain sensitivities on the blade surface, an additional adjoint mesh deformation has to be performed, as detailed in the publication of Engels-Putzka and Backhaus \cite{EB}.

\begin{figure}[b!]
	\vspace{-10pt}
	\centering
	\includegraphics[width=1.0\linewidth]{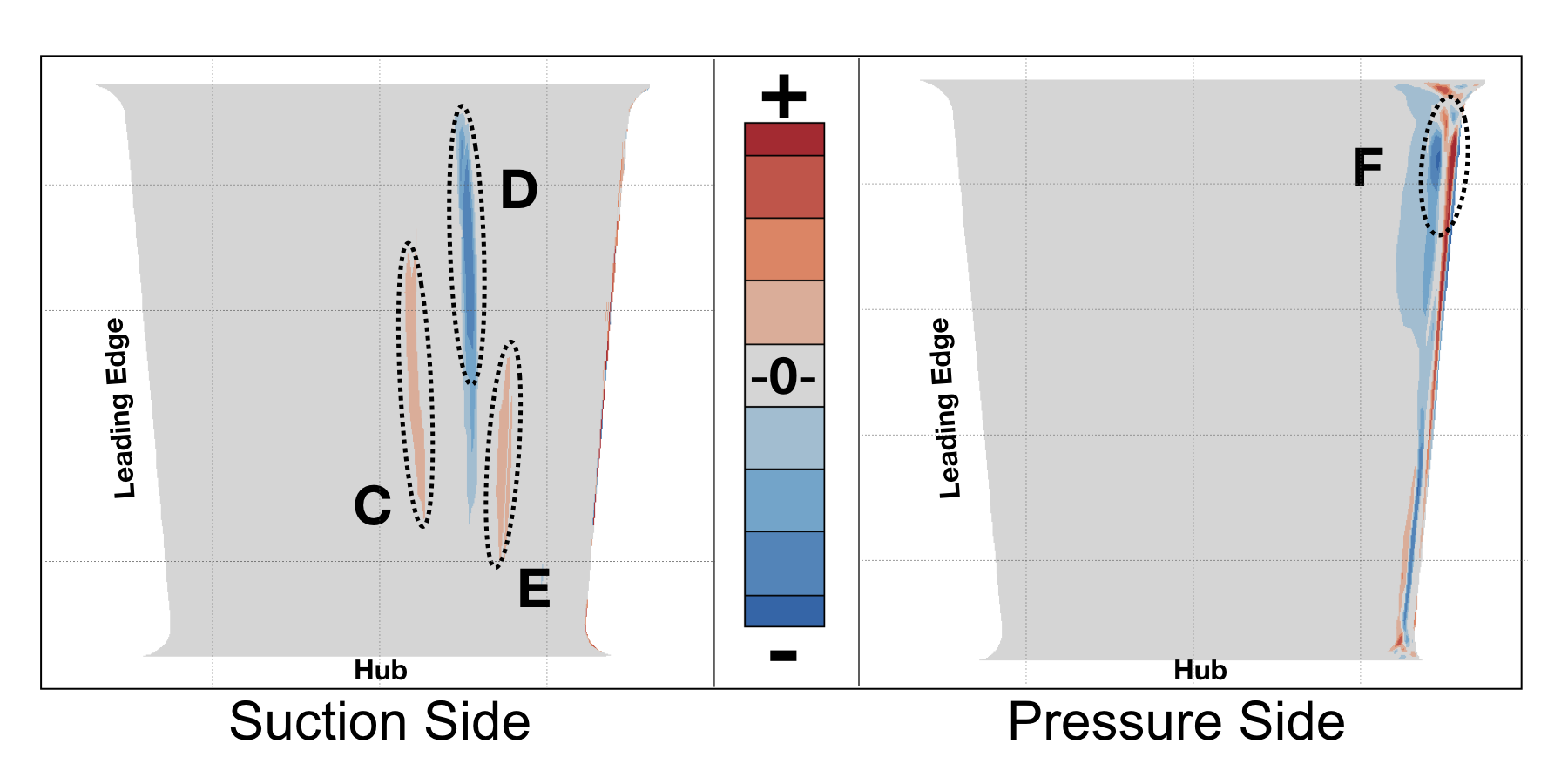}
	\vspace{-20pt}
	\caption{Pressure loss coefficient surface sensitivities. Scaled suction and pressure side. Red/blue color highlighting surface outward/inward movement in normal direction. Plus/minus sign indicating increase/decrease in pressure loss coefficient}
	\label{fig:surfaceSensePressureLoss}
	\vspace{-10pt}
\end{figure}
Figure \ref{fig:surfaceSenseMassFlow} shows the sensitivity map for the mass flow objective on the suction and pressure side. A high negative sensitivity can be seen on the suction side as highlighted by zone A. This indicates that an inward movement of the blade surface would increase the mass flow. Zone A coincides with the region at which the flow reaches Mach 1 and an increase in blade throat area would therefore benefit the mass flow.
The highest negative sensitivity for the mass flow is on the pressure side near the trailing edge (TE), showing a switch between inward to outward movement of the TE blade surface. The surface inward movement in zone B would prevent the flow from entering a transonic regime and thus would allow an increased mass flow. The outward movement in zone B near the TE shows a high decrease in mass flow. This might be achieved through a reduction in TE thickness by creating a sharper and longer TE.

The sensitivity of the pressure loss coefficient is displayed in Figure \ref{fig:surfaceSensePressureLoss}. The areas of highest sensitivity are on the suction side  zone C, D and E. Zone C and E highlight areas near Mach 1. An outward surface movement in zone C and E would increase the pressure loss, by increasing the surface curvature on the suction side, leading to a higher Mach Number and thus more loss. Contrary, an inward movement of zone D would decrease the pressure loss over the vane. A lower Mach number would lead to less shock losses and therefore to less pressure loss.
On the pressure side of the turbine vane the highest sensitivity for the pressure loss coefficient is near the upper TE, as highlighted by zone F. An inward movement would reduce the surface curvature in area F and would thus lead to less losses. The outward movement near the TE shows that the pressure loss would increase, which might be caused by the increase in TE thickness.

\subsubsection{Impact of Manufacturing Variations}
\begin{figure}[b]
	\begin{minipage}{0.49\textwidth}
		\centering
		\includegraphics[width=\linewidth]{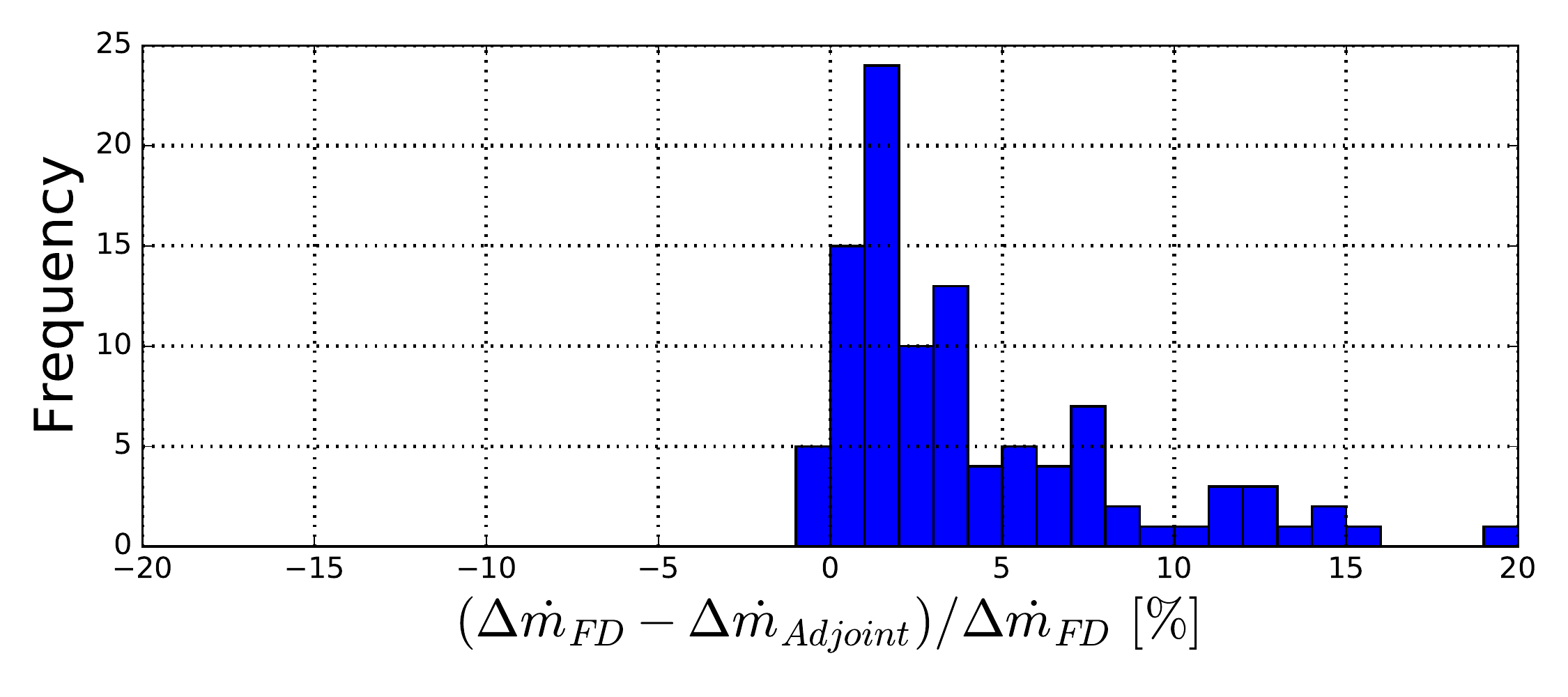}
		\subcaption{Mass flow objective}
		\label{fig:histogram_massflow}
	\end{minipage}
	\begin{minipage}{0.49\textwidth}
		\centering
		\includegraphics[width=\linewidth]{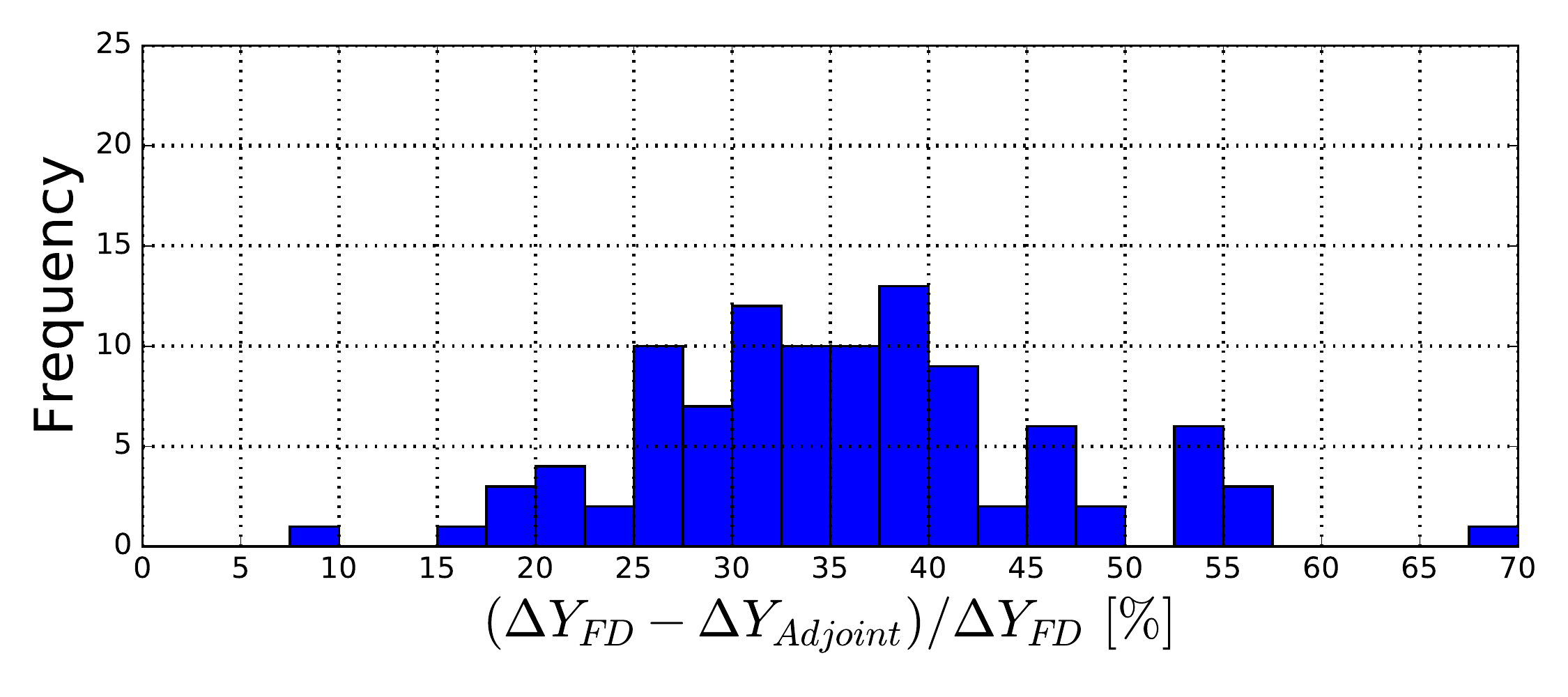}
		\subcaption{Pressure loss coefficient objective}
		\label{fig:histogram_pressureloss}
	\end{minipage}
	\caption{Adjoint gradient deviations in regard to finite differences}
\end{figure}
\begin{figure}[t]
	\begin{minipage}{0.49\textwidth}
		\centering
		\includegraphics[width=\linewidth]{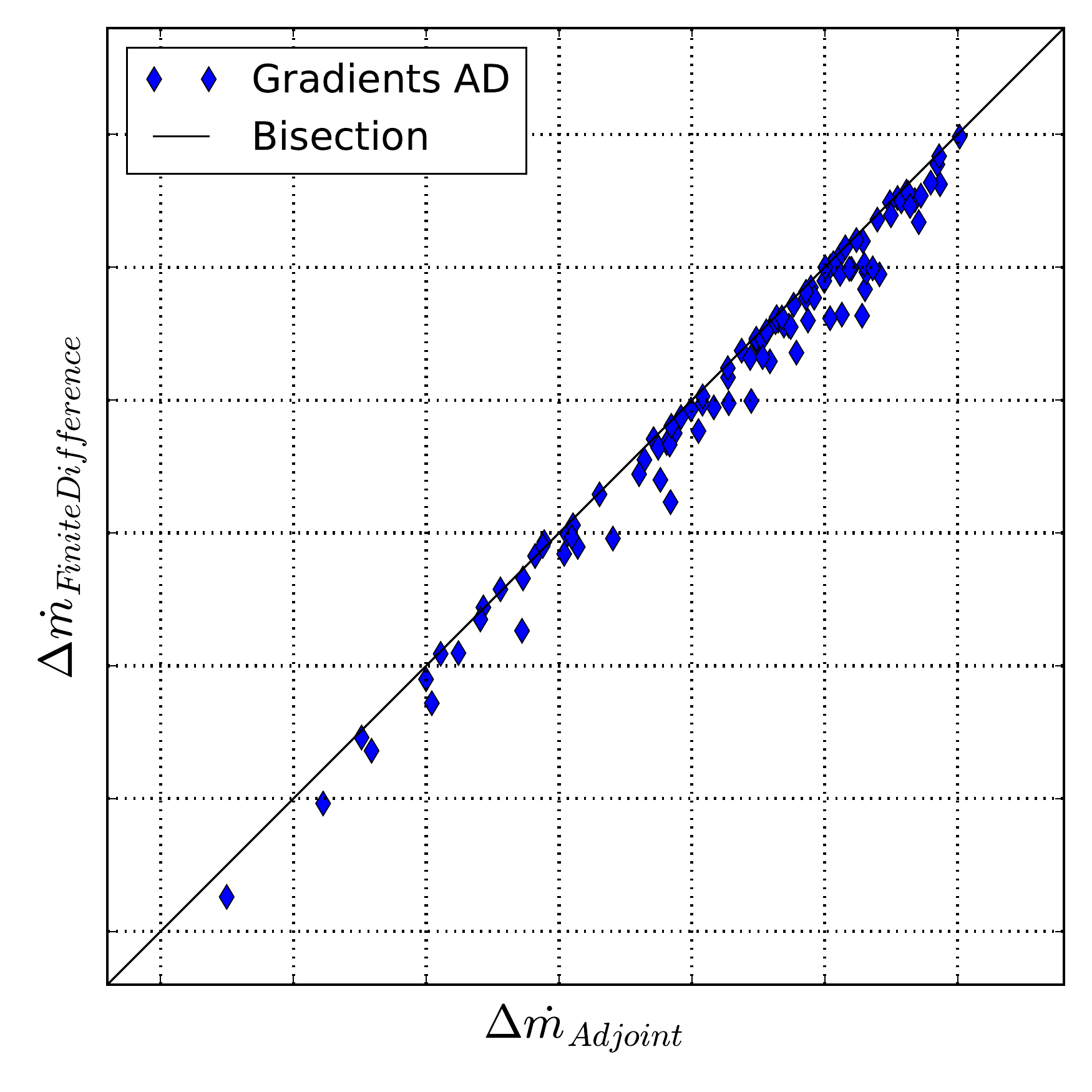}
		\subcaption{Mass flow objective}
		\label{fig:bisection_massflow}
	\end{minipage}
	\hfill
	\begin{minipage}{0.49\textwidth}
		\centering
		\includegraphics[width=\linewidth]{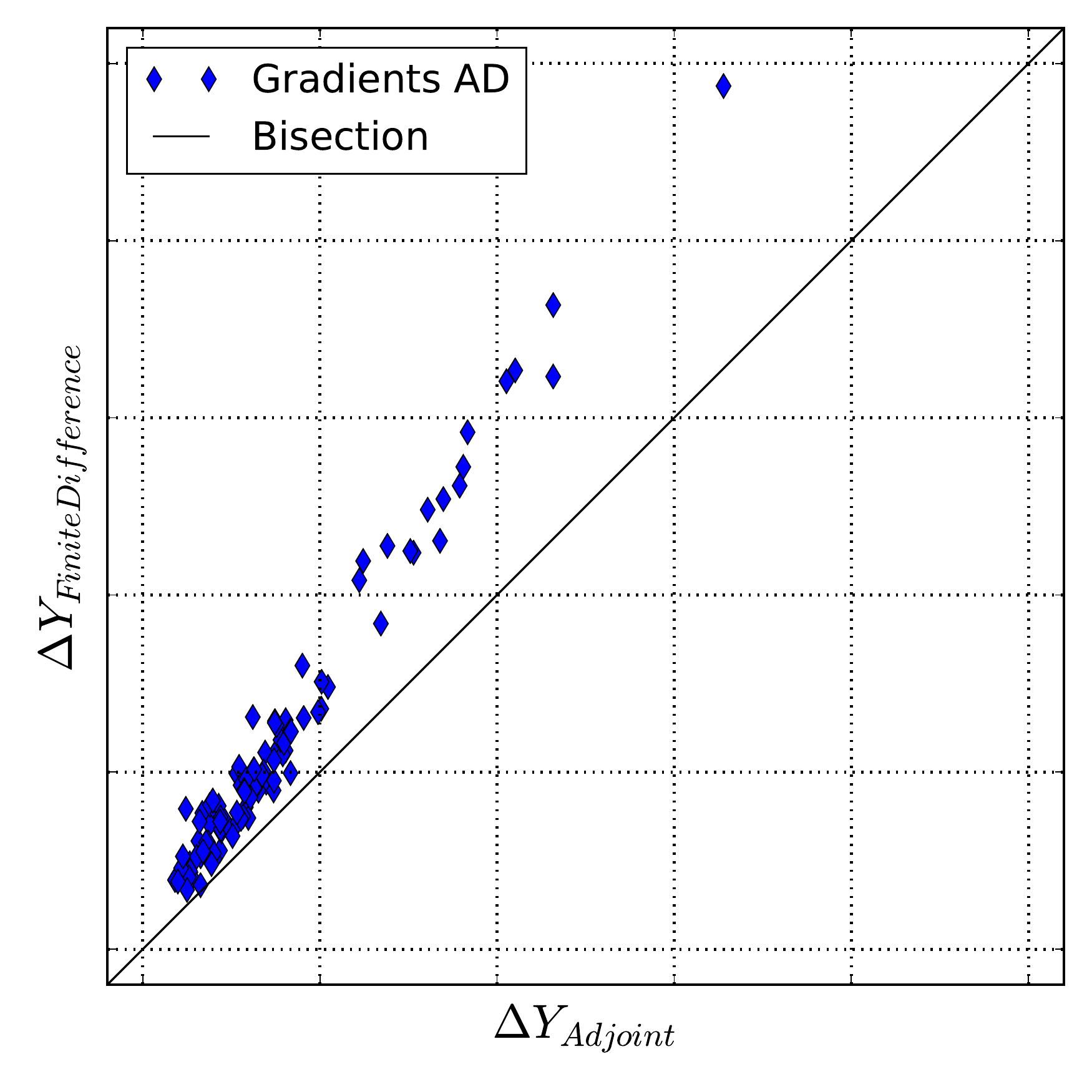}
		\subcaption{Pressure loss coefficient objective}
		\label{fig:bisection_pressureloss}
	\end{minipage}
	\caption{Adjoint gradients versus finite difference gradients}
	\vspace{-10pt}
\end{figure}

The  mass flow gradient deviation histogram for all 102 deformed meshes is shown in Figure \ref{fig:histogram_massflow}. For the mass flow objective almost all deviations are positive, indicating that the adjoint gradients underestimated the change in mass flow. In total around 90\% of all meshes have a gradient deviation of less than +10\%, while only one deformed mesh has a maximum deviation of +20\%. 
Compared to the mass flow objective, the pressure loss coefficient gradient deviations range from +10\% to 70\% as displayed in Figure \ref{fig:histogram_pressureloss}. The average gradient deviation is around 35\%.

The gradient deviations for the mass flow agree very well with the previous results from the mass flow parameter study, which were around 2\%. Considering that in this case the complete blade surface is varied, an increased gradient deviation is acceptable. However, the high average deviation for the pressure loss coefficient indicates a problem with the implemented adjoint objective or a nonlinear flow response in regard to the pressure loss objective.

\subsubsection*{Linear Flow Response}
To investigate if the impact of the manufacturing variations may be regarded as linear, the finite difference and adjoint gradients are plotted against each other for both objectives as shown in Figure \ref{fig:bisection_massflow} and \ref{fig:bisection_pressureloss}. Additionally, the bisection is also displayed, proving the linear relationship between the gradient couples.

The mass flow finite difference gradients clearly agree with the adjoint gradient results, demonstrating that the mapped MVs can be evaluated by the linear adjoint approach. The gradients for the pressure loss coefficient also show a linear behavior as displayed in Figure \ref{fig:bisection_pressureloss}. 

The computed adjoint gradients, however, deviate around 35\% compared to the finite differences. This almost constant deviation is caused by a geometrical artifact - independent of adjoint computation - introduced during the mesh morphing. As previously mentioned, the mesh morphing is only applied between 1 and 99\% span, which leads to a step-like edge at the transition point. The step height is dependent on the magnitude of the MVs and leads to an additional pressure loss caused by nonlinear flow phenomena, resulting in the adjoint gradient deviation. This problem will be addressed in a following publication.  

Despite this slight systematic deviation for one of the objectives, the adjoint method correctly predicts sign and magnitude for all 102 deformed meshes. Thereby proving that the applied MVs are small enough to be modeled by the adjoint method.

\subsubsection{Computational Efficiency of the Adjoint-based Method}
One of the main limitations of using RANS computations to evaluate the impact of MVs is the  linear dependency on the number of parameters and the computational cost in run time.
The adjoint approach, however, is independent of the number of parameters and only requires an additional adjoint computations for each investigated objective.

Sagebaum et al. \cite{SOGBFMN} measure the computational cost for the AdjointTrace AD solver with an increased CPU time factor of 1.5 and a memory factor of 8.3 compared to the primal CFD evaluation. 
For the turbine vane considered in this work, these figures are summarized in Table \ref{tab:vaneRunTimeNew}. It can be noticed that the performance registered for this test case is very close to the observation made by Sagebaum et al. \cite{SOGBFMN}.
An adjoint evaluation requires 30\% more CPU time and 8.8 times more memory. The quantitative increase in memory is mainly caused by the requirement to tape one primal iteration step of the adjoint iteration.

Nevertheless, the main advantage is the reduction in run time as shown in Table \ref{tab:vaneRunTime2New}. The computation of 102 RANS evaluations takes 28 times longer,  compared to one RANS and two adjoint evaluations. 

\begin{table}[t]
	\begin{minipage}{.4\linewidth}
		\centering
		\caption{Numerical cost of adjoint}
		\begin{tabular}{|l|c|c|}
			\hline
			& \textbf{CPU} & \textbf{RAM} \\ 
			\hline
			\textbf{Primal} & 1 & 1 \\ 
			\hline 
			\textbf{Adjoint} & 1.3 & 8.8 \\ 
			\hline			
		\end{tabular} 

		\label{tab:vaneRunTimeNew}
	\end{minipage}
	\hfill
	\begin{minipage}{.6\linewidth}
		\centering
		\caption{Computational time benefit}
		\begin{tabular}{|l||c|c||c|}
			\hline
			\textbf{Approach} & \textbf{RANS} & \textbf{Adjoint} & \textbf{Time} \\ 
			\hline 
			\textbf{Nonlinear} & 102 & - & 100\% \\ 
			\hline 
			\textbf{Adjoint} & 1 & 2 & 3.5\% \\ 
			\hline
		\end{tabular} 	
		\label{tab:vaneRunTime2New}
	\end{minipage}
\end{table}

\section{Conclusions}

In this paper, an adjoint-based toolchain has been validated to quantify the impact of MVs on aerodynamic performance. Two different adjoint versions, a hand-derived and algorithmic-derived, of the CFD suite Trace were compared with each other in a systematic parameter study.
The adjoint gradients were validated against finite differences, showing that the gradient accuracy is dependent on the objective and parameter, as well as that the algorithmically-differentiated version is more accurate.

The adjoint-based toolchain was then applied to an industrial heavy duty turbine vane. The MVs were derived from 102 optical white light scans, using a mesh morphing approach to directly impose the MVs on the baseline geometry.
The validation shows that the adjoint method produces results of similar accuracy as RANS evaluations. Furthermore, the results prove that the impact of MVs on mass flow and pressure loss is small enough to be considered linear.

As shown, the adjoint-based approach is an alternative to computational expensive RANS computations for evaluating the impact of MVs. 
A reduction by a factor of 28 was achieved by applying the adjoint method, providing a fast and accurate tool for blade design.

\section*{Acknowledgment}
The permission of Siemens AG to publish the results is greatly acknowledged. The author would also like to thank the DLR, Institute of Propulsion Technology, for their support with Trace, especially Jan Backhaus, as well as Johannes Steiner from Siemens for the discussions and support regarding the manufacturing variation analysis. H. G. gratefully acknowledges partial financial support by the Federal Ministry of Research and Education BMBF via the GIVEN project, grant no. 05M18PXA.


\begin{thebibliography}{123}
\bibitem{GD} Garzon, V. E. and Darmofal, D. L. “Impact of geometric variability on axial compressor performance”. ASME Turbo Expo 2003: Power for Land, Sea, and Air. American Society of Mechanical
Engineers. 2003.
\bibitem{Duf} Duffner, J. D. “The effects of manufacturing variability on turbine vane performance”. MA thesis.
Massachusetts Institute of Technology, 2008.
\bibitem{LVVSJG} Lange, A., Voigt, M., Vogeler, K., Schrapp, H., Johann, E., and Gummer, V. “Probabilistic CFD
simulation of a high-pressure compressor stage taking manufacturing variability into account”.
ASME Turbo Expo 2010: Power for Land, Sea, and Air. American Society of Mechanical Engineers.
2010.
\bibitem{GBFS} Giebmanns, A., Backhaus, J., Frey, C., and Schnell, R. “Compressor leading edge sensitivities and
analysis with an adjoint flow solver”. ASME Turbo Expo 2013: Turbine Technical Conference and
Exposition. American Society of Mechanical Engineers. 2013.
\bibitem{ZBB} Zamboni, G., Banks, G., and Bather, S. “Gradient-based adjoint and design of experiment cfd
methodologies to improve the manufacturability of high pressure turbine blades”. ASME Turbo
Expo 2016: Turbine Technical Conference and Exposition. 2016.
\bibitem{YXMHLL} Yang, J., Xiong, J., McBean, I., Havakechian, S., Liu, F., and Luo, J. “Performance impact of
manufacturing variations for multistage steam turbines”. Journal of Propulsion and Power (2017).
\bibitem{Gal} Gallus, H. “ERCOFTAC test case 6: axial flow turbine stage”. Seminar and Workshop on 3D
Turbomachinery flow prediction III, Les Arcs, France. 1995.
\bibitem{VFK} Voigt, C., Frey, C., and Kersken, H. “Development of a generic surface mapping algorithm for fluid-
structure-interaction simulations in turbomachinery”. 5th European Conference on Computational
Fluid Dynamics (ECCOMAS CFD 2010). 2010.
\bibitem{FKN} Frey, C., Kersken, H.-P., and N\"urnberger, D. “The discrete adjoint of a turbomachinery RANS
solver”. ASME Turbo Expo 2009: Power for Land, Sea, and Air. American Society of Mechanical
Engineers. 2009.
\bibitem{FABKW} Frey, C., Ashcroft, G., Backhaus, J., Kugeler, E., and Wellner, J. “Adjoint-based flow sensitivity
analysis using arbitrary control surfaces”. ASME 2011 Turbo Expo: Turbine Technical Conference
and Exposition. American Society of Mechanical Engineers. 2011.
\bibitem{BSFMNSG} Backhaus, J., Schmitz, A., Frey, C., Mann, S., Nagel, M., Sagebaum, M., and Gauger, N. R.
“Application of an algorithmically differentiated turbomachinery flow solver to the optimization
of a fan stage”. 18th AIAA/ISSMO Multidisciplinary Analysis and Optimization Conference. 2017.
\bibitem{SOGBFMN} Sagebaum, M., Ozkaya, E., Gauger, N. R., Backhaus, J., Frey, C., Mann, S., and Nagel, M.
“Efficient algorithmic differentiation techniques for turbo-machinery design”. 18th AIAA/ISSMO
Multidisciplinary Analysis and Optimization Conference. 2017.
\bibitem{SHVVM} Scharfenstein, J., Heinze, K., Voigt, M., Vogeler, K., and Meyer, M. “Probabilistic CFD analysis of
high pressure turbine blades considering real geometric effects”. ASME Turbo Expo 2013: Turbine
Technical Conference and Exposition. American Society of Mechanical Engineers. 2013.
\bibitem{RJGG} Restemeier, M., Jeschke, P., Guendogdu, Y, and Gier, J. “Numerical and experimental analysis of
the effect of variable blade row spacing in a subsonic axial turbine”. Journal of Turbomachinery
(2013).
\bibitem{Wil} Wilcox, D. C. “Reassessment of the scale-determining equation for advanced turbulence models”.
AIAA journal 26.11 (1988), pp. 1299–1310.
\bibitem{Kat} Kato, M. “The modeling of turbulent flow around stationary and vibrating square cylinders”.
Ninth Symposium on Turbulent Shear Flows, 1993. 1993.
\bibitem{EB} Engels-Putzka, A. and Backhaus, J. “Adjoint-based Shape Sensitivities for Turbomachinery Design
Optimizations”. EUROGEN (2013).
\end{thebibliography}
\end{document}